\input harvmac.tex

\input epsf.tex

\def\figin{\epsfcheck\figin}\def\figins{\epsfcheck\figins}
\def\epsfcheck{\ifx\epsfbox\UnDeFiNeD
\message{(NO epsf.tex, FIGURES WILL BE IGNORED)}
\gdef\figin##1{\vskip2in}\gdef\figins##1{\hskip.5in}
\else\message{(FIGURES WILL BE INCLUDED)}%
\gdef\figin##1{##1}\gdef\figins##1{##1}\fi}
\def\DefWarn#1{}
\def\figinsert{\goodbreak\midinsert}
\def\ifig#1#2#3{\DefWarn#1\xdef#1{fig.~\the\figno}
\writedef{#1\leftbracket fig.\noexpand~\the\figno}%
\figinsert\figin{\centerline{#3}}\medskip\centerline{\vbox{\baselineskip12pt
\advance\hsize by -1truein\noindent\footnotefont{\bf Fig.~\the\figno:} #2}}
\bigskip\endinsert\global\advance\figno by1}




{ \Title{
\vbox{\baselineskip12pt 
 }}
 {\vbox{
{\centerline { Drag force in a string}
{\centerline { model dual to large-$N$ QCD} }
}}}}

\bigskip
\centerline{ Pere Talavera }
\bigskip~
 



\medskip

\centerline{Departament de F{\'\i}sica i Enginyeria Nuclear,}
\centerline{  Universitat Polit\`ecnica de Catalunya, Jordi Girona 1-3, E-08034 Barcelona, Spain}


\medskip

\vskip .3in

\baselineskip12pt

\vfill
We compute the drag force exerted on a quark and a di-quark systems in a background dual to large-N QCD  at finite temperature. We find that appears a drag force in the former setup  with 
flow of energy proportional to the mass of the quark while in the latter there is no dragging as in other studies. We also review the screening length.

\Date{October 2006} \eject \baselineskip14pt

\lref\WittenZW{
  E.~Witten,
   ``Anti-de Sitter space, thermal phase transition, and confinement in  gauge
  theories,''
  Adv.\ Theor.\ Math.\ Phys.\  {\bf 2}, 505 (1998)
  [arXiv:hep-th/9803131].
}

\lref\AharonyBM{
  O.~Aharony, S.~Minwalla and T.~Wiseman,
  ``Plasma-balls in large N gauge theories and localized black holes,''
  Class.\ Quant.\ Grav.\  {\bf 23}, 2171 (2006)
  [arXiv:hep-th/0507219].
}

\lref\FriessAW{
  J.~J.~Friess, S.~S.~Gubser and G.~Michalogiorgakis,
   ``Dissipation from a heavy quark moving through N = 4 super-Yang-Mills
  plasma,''
  arXiv:hep-th/0605292.
}

\lref\GubserBZ{
  S.~S.~Gubser,
  ``Drag force in AdS/CFT,''
  arXiv:hep-th/0605182.
}

\lref\HerzogGH{
  C.~P.~Herzog, A.~Karch, P.~Kovtun, C.~Kozcaz and L.~G.~Yaffe,
   ``Energy loss of a heavy quark moving through N = 4 supersymmetric Yang-Mills
  plasma,''
  JHEP {\bf 0607}, 013 (2006)
  [arXiv:hep-th/0605158].
}

\lref\CaceresAS{
  E.~Caceres and A.~Guijosa,
  ``On drag forces and jet quenching in strongly coupled plasmas,''
  arXiv:hep-th/0606134.
}

\lref\HuotYS{
  S.~C.~Huot, S.~Jeon and G.~D.~Moore,
   ``Shear viscosity in weakly coupled N = 4 super Yang-Mills theory compared to
  QCD,''
  arXiv:hep-ph/0608062.
}

\lref\KarchSH{
  A.~Karch and E.~Katz,
  ``Adding flavor to AdS/CFT,''
  JHEP {\bf 0206}, 043 (2002)
  [arXiv:hep-th/0205236].
}

\lref\NeriIC{
  F.~Neri and A.~Gocksch,
  ``Chiral Symmetry Restoration In Large N QCD At Finite Temperature,''
  Phys.\ Rev.\ D {\bf 28}, 3147 (1983).
}

\lref\PisarskiDB{
  R.~D.~Pisarski,
  ``Finite Temperature QCD At Large N,''
  Phys.\ Rev.\ D {\bf 29}, 1222 (1984).
}

\lref\PeetersIU{
  K.~Peeters, J.~Sonnenschein and M.~Zamaklar,
   ``Holographic melting and related properties of mesons in a quark gluon
  plasma,''
  arXiv:hep-th/0606195.
}

\lref\MaldacenaIM{
  J.~M.~Maldacena,
  ``Wilson loops in large N field theories,''
  Phys.\ Rev.\ Lett.\  {\bf 80}, 4859 (1998)
  [arXiv:hep-th/9803002].
}

\lref\BrandhuberER{
  A.~Brandhuber, N.~Itzhaki, J.~Sonnenschein and S.~Yankielowicz,
   ``Wilson loops, confinement, and phase transitions in large N gauge  theories
  from supergravity,''
  JHEP {\bf 9806}, 001 (1998)
  [arXiv:hep-th/9803263].
}

\lref\BrandhuberBS{
  A.~Brandhuber, N.~Itzhaki, J.~Sonnenschein and S.~Yankielowicz,
  ``Wilson loops in the large N limit at finite temperature,''
  Phys.\ Lett.\ B {\bf 434}, 36 (1998)
  [arXiv:hep-th/9803137].
}

\lref\ReyBQ{
  S.~J.~Rey, S.~Theisen and J.~T.~Yee,
   ``Wilson-Polyakov loop at finite temperature in large N gauge theory and
  anti-de Sitter supergravity,''
  Nucl.\ Phys.\ B {\bf 527}, 171 (1998)
  [arXiv:hep-th/9803135].
}

\lref\ReyIK{
  S.~J.~Rey and J.~T.~Yee,
   ``Macroscopic strings as heavy quarks in large N gauge theory and  anti-de
  Sitter supergravity,''
  Eur.\ Phys.\ J.\ C {\bf 22}, 379 (2001)
  [arXiv:hep-th/9803001].
}

\lref\ChernicoffHI{
  M.~Chernicoff, J.~A.~Garcia and A.~Guijosa,
  ``The energy of a moving quark-antiquark pair in an N = 4 SYM plasma,''
  arXiv:hep-th/0607089.
}

\lref\ParnachevDN{
  A.~Parnachev and D.~A.~Sahakyan,
  ``Chiral phase transition from string theory,''
  arXiv:hep-th/0604173.
}

\lref\AharonyDA{
  O.~Aharony, J.~Sonnenschein and S.~Yankielowicz,
  ``A holographic model of deconfinement and chiral symmetry restoration,''
  arXiv:hep-th/0604161.
}

\lref\ArgyresVS{
  P.~C.~Argyres, M.~Edalati and J.~F.~Vazquez-Poritz,
   ``No-drag string configurations for steadily moving quark-antiquark pairs in
  a thermal bath,''
  arXiv:hep-th/0608118.
}

\lref\AvramisEM{
  S.~D.~Avramis, K.~Sfetsos and D.~Zoakos,
   ``On the velocity and chemical-potential dependence of the heavy-quark
  interaction in N = 4 SYM plasmas,''
  arXiv:hep-th/0609079.
}

\lref\ArseneFA{
  I.~Arsene {\it et al.}  [BRAHMS Collaboration],
   ``Quark gluon plasma and color glass condensate at RHIC? The perspective
  from the BRAHMS experiment,''
  Nucl.\ Phys.\ A {\bf 757}, 1 (2005)
  [arXiv:nucl-ex/0410020].
}

\lref\AdcoxMH{
  K.~Adcox {\it et al.}  [PHENIX Collaboration],
   ``Formation of dense partonic matter in relativistic nucleus nucleus
  collisions at RHIC: Experimental evaluation by the PHENIX  collaboration,''
  Nucl.\ Phys.\ A {\bf 757}, 184 (2005)
  [arXiv:nucl-ex/0410003].
}

\lref\BackJE{
  B.~B.~Back {\it et al.},
  ``The PHOBOS perspective on discoveries at RHIC,''
  Nucl.\ Phys.\ A {\bf 757}, 28 (2005)
  [arXiv:nucl-ex/0410022].
}

\lref\AdamsDQ{
  J.~Adams {\it et al.}  [STAR Collaboration],
   ``Experimental and theoretical challenges in the search for the quark  gluon
   plasma: The STAR collaboration's critical assessment of the  evidence from
  RHIC collisions,''
  Nucl.\ Phys.\ A {\bf 757}, 102 (2005)
  [arXiv:nucl-ex/0501009].
}

\lref\GaoSE{
  Y.~h.~Gao, W.~s.~Xu and D.~f.~Zeng,
  ``Wake of color fileds in charged N = 4 SYM plasmas,''
  arXiv:hep-th/0606266.
}

\lref\LiuUG{
  H.~Liu, K.~Rajagopal and U.~A.~Wiedemann,
  ``Calculating the jet quenching parameter from AdS/CFT,''
  arXiv:hep-ph/0605178.
}

\lref\BuchelBV{
  A.~Buchel,
   ``On jet quenching parameters in strongly coupled non-conformal gauge
  theories,''
  Phys.\ Rev.\ D {\bf 74}, 046006 (2006)
  [arXiv:hep-th/0605178].
}

\lref\VazquezPoritzBA{
  J.~F.~Vazquez-Poritz,
  ``Enhancing the jet quenching parameter from marginal deformations,''
  arXiv:hep-th/0605296.
}

\lref\ArmestoZV{
  N.~Armesto, J.~D.~Edelstein and J.~Mas,
   ``Jet quenching at finite 't Hooft coupling and chemical potential from
  AdS/CFT,''
  arXiv:hep-ph/0606245.
}

\lref\LinAU{
  F.~L.~Lin and T.~Matsuo,
  ``Jet quenching parameter in medium with chemical potential from AdS/CFT,''
  Phys.\ Lett.\ B {\bf 641}, 45 (2006)
  [arXiv:hep-th/0606136].
}

\lref\ShuryakIA{
  E.~Shuryak, S.~J.~Sin and I.~Zahed,
  ``A gravity dual of RHIC collisions,''
  arXiv:hep-th/0511199.
}

\lref\FriessRK{
  J.~J.~Friess, S.~S.~Gubser, G.~Michalogiorgakis and S.~S.~Pufu,
  ``Stability of strings binding heavy-quark mesons,''
  arXiv:hep-th/0609137.
}

\lref\bj{
  J.~D.~Bjorken,
   ``Highly Relativistic Nucleus-Nucleus Collisions: The Central Rapidity
  Region,''
  Phys.\ Rev.\ D {\bf 27}, 140 (1983).
}

\lref\baier{  R.~Baier, D.~Schiff and B.~G.~Zakharov,
  ``Energy loss in perturbative QCD,''
  Ann.\ Rev.\ Nucl.\ Part.\ Sci.\  {\bf 50}, 37 (2000)
  [arXiv:hep-ph/0002198].
}

\lref\CaceresTA{
  E.~Caceres, M.~Natsuume and T.~Okamura,
  ``Screening length in plasma winds,''
  arXiv:hep-th/0607233.
}

\lref\CaceresDJ{
  E.~Caceres and A.~Guijosa,
  ``Drag force in charged N = 4 SYM plasma,''
  arXiv:hep-th/0605235.
}

\lref\HerzogSE{
  C.~P.~Herzog,
  ``Energy loss of heavy quarks from asymptotically AdS geometries,''
  JHEP {\bf 0609}, 032 (2006)
  [arXiv:hep-th/0605191].
}

\newsec{Motivation and results}

QCD at $T\neq 0$ plays an important role in understanding two related areas in physics: the physics of the early universe and the physics of heavy ions collisions. Due to the increasing experimental evidence for the existence of the quark-gluon plasma (QGP) state at RHIC \refs{\ArseneFA, \AdcoxMH,\BackJE,\AdamsDQ} and the existing planned experiments as LHC and FAIR we shall be concerned with the latter.  The experimental results are against the old prejudices that assumed that at sufficient high temperatures the interactions between quarks and gluons would lie in the perturbative regime making of the QGP an almost ideal gas. The data suggest that in a QGP matter is almost, but totally, in the deconfined phase but with the  't Hooft coupling constant in the ballpark of a decade invalidating any perturbative approach, $g^2_{\rm YM} N \sim 10.$ This makes of QGP a good testing ground of the AdS/CFT duality that 
has as a prerequisite the condition $g^2_{\rm YM} N >> 1.$ In particular, to model this physics, we shall use a setup of D4 branes at finite temperature \AharonyBM~ but with a zero chemical potential.  This, in principle, does not present an obstacle to attain realistic results inasmuch we are just interested in the fluid properties produced in the central rapidity region \bj.

We shall be concerned mainly with one measured quantity, the energy lost of a high-energy parton in the QGP \baier. We look at two different systems: a single heavy-quark and a di-quark  passing through the QGP. To make connection with string theory one models the single external quark as a string dangling from the boundary and the di-quark system as a string hanging between the two quarks. Both setups have been extensively studied recently in the same context but with different backgrounds \refs{ \HerzogGH, \GubserBZ, \LiuUG, \HerzogSE, \CaceresDJ, \FriessAW, \CaceresAS,\GaoSE,\BuchelBV,\VazquezPoritzBA,\LinAU,\ShuryakIA,\ArmestoZV}. These strings extend on the radial transverse coordinate and can be thought as a flux tube spreading out of the $3+1$ boundary theory. This flux tube will travel along the bulk that is constituted by a thermal bath of gluons. The interaction with this gas will exert a force on the string that is propagated along it to the external quarks at the boundary. This is the effect we are precisely interested in.

As we shall explicitly see there are no qualitative changes in our results to what is already known in the literature: a single quark  experiences a drag force while in the di-quark system there is no dragging in the direction of the movement. These seem to be the general pattern to all the models studied so far.

The paper is organized as follows:  sec. 2 contains the two supergravity backgrounds dual to large-N QCD we are considering, one at zero temperature and the equivalent at finite temperature. In sec. 3  we study, in both phases, the
force experienced by a heavy-quark passing through a plasma and discuss the thermal mass. Sec. 4 contains the study, in the decofined regime, of a pair of quarks. We focus the attention in the  energy-distance function. Because is not clear so far whether from this or similar studies can be obtained the jet-quenching parameter itself, $\hat{q}$, we retreat to go further and obtain it from our results. Furthermore an analysis on the stability \FriessRK~or favored energetic configuration \ArgyresVS~is lacking in our study.

\newsec{Zero vs. finite temperature QCD at large-N}

We briefly review the two backgrounds we used. In doing so we must bear in mind that the field theory we expect to describe is at the boundary of both backgrounds and therefore they must asymptote one to each other in this limit.
The reason to use two different setups is related to the confined-deconfined  phase transition. At zero chemical potential, we expect a cross-over region between these two phases in which hadrons behave so differently. Notice, that we do not use probe-branes in our analysis and hence, we are mainly testing properties of the thermal bath while matter is treated as a external source.

At zero temperature we use a stack of D4-branes (both models are in Lorentian signature) were the initial time coordinate has been compactified on a circle of radii $u_\Lambda$. This breaks supersymmetry and give masses to the fermions at one loop. After this we Wick rotate one of the remaining 
space coordinates that will play the role of time. The final form of the various fields: metric, the RR four-form and the dilaton, are \WittenZW
$$
ds^2= \left({u\over R}\right)^{3/2} \left[ -dt^2 + \delta_{ij} dx^i dx^j + f(u) dx_4^2 \right] +\left({u\over R}\right)^{-3/2} \left[ {du^2\over f(u)} + u^2 d\Omega_4^2 \right]\,,
$$
\eqn\zero{  F_{(4)}={2\pi N_c\over V_4}\epsilon_4 \,,\quad  e^\phi=g_s \left({u\over R}\right)^{3/4}\,, \quad R^3 =\pi g_s
N_c l_s^3\,, \quad f(u)= 1 - \left({u_\Lambda\over u}\right)^3\,.
}
The surface spanned by $x_4-u$ has a cigar shape that is not singular only if
$$\delta x_4 = {4\pi\over 3} \left({R^3\over u_\Lambda} \right)^{1/2} .$$
Notice that as the temporal coordinate can not develop a singularity after compactification the horizon coordinate is not related with the temperature of the field theory, and we are free to choose $T=0$.
The peculiarities of \zero~ are already well studied in the literature: {\sl i)} It confines. {\sl ii)} It has chiral symmetry breaking when probe-branes are added.  {\sl iii)} There are indications for the decoupling of the Kaluza-Klein modes from the spectrum. 

The second of our backgrounds is an euclidean continuation of \zero~ and is assumed to describe physics above the critical region, $T>T_c$: time is compactified over a circle of period $\beta$. Anti-periodic boundary conditions for the fermionic content of the theory is imposed along $t$ and $x_4$-directions. The latter with radii $2\pi R$. 

The field content is
$$
ds^2= \left({u\over R}\right)^{3/2} \left[- f(u) dt^2 + \delta_{ij} dx^i dx^j + dx_4^2 \right] +\left({u\over R}\right)^{-3/2} \left[  {du^2\over f(u)} + u^2 d\Omega_4^2 \right]\,,
$$
\eqn\finite{  F_{(4)}={2\pi N_c\over V_4}\epsilon_4 \,,\quad  e^\phi=g_s \left({u\over R}\right)^{3/4}\,, \quad R^3 =\pi g_s
N_c l_s^3\,, \quad f(u)= 1 - \left({u_T\over u}\right)^3\,,
}
where to avoid time direction singularities one must choose
$$ \delta t = {4\pi\over 3} \left({R^3\over u_T} \right)^{1/2} =\beta\,.$$

Because there is no smooth interpolation between the two backgrounds, \zero-\finite, the system is believed to describe a first order transition. 
By looking at the free energy one sees that \zero~is the relevant background at zero temperature while \finite~is at high ones \refs{\AharonyBM,\ParnachevDN}.  The relevant salient point of \finite~is that while it supports a discrete hadron spectrum, chiral symmetry breaking is still at work, it has occurred deconfinement, the gluonic sector has lost its symmetry. Thus all in all, \finite~can contain quarks, gluons and hadrons simultaneously as degrees of freedom.   This picture, although not very intuitive, is in agreement with the QGP data where at $T=T_c$ not all the hadronic dof are lost \ArseneFA.

\newsec{Drag force for a single quark}

We focus first in the study of the force exerted by an external agent on a quark \refs{\HerzogGH, \GubserBZ}.
For that purpose we fixed all but the five uncompact dimensions, from which all transverse coordinates
vanish except one $x$. Furthermore we work on the static gauge $t=\tau, u=\sigma, x(\tau,\sigma)$. The action of a fundamental string
in the string frame is given by the Nambu-Goto action
\eqn\nambugoto{ S =  - {1\over 2 \pi \alpha} \int d\sigma d\tau \sqrt{-g}\,.}

\subsec{Deconfined phase}

With the previous, static gauge, ansatz  the lagrangian density for \finite~ reads
\eqn\lags{ {\cal L} = -{1\over 2 \pi \alpha} \sqrt{ 1-  {\left( \dot{x}\right)^2  \over f }+ h f \left(x^\prime\right)^2} 
\,,}
where dots (primes) stand for derivatives w.r.t. $\tau~ (\sigma)\,.$ We have defined $h\equiv  \left({u\over R}\right)^{3}$ and dropped all the arguments in the functions. The e.o.m. for $x$ is
\eqn\eomcon{ 
 \partial_\tau\left({\dot{x}\over \sqrt{-g} }\right) - f
\partial_\sigma \left( {f h x^\prime \over \sqrt{-g} }\right) = 0\,.
}
Notice that the solution of a static string, $x=ct$, stretching from the horizon up to the boundary exists.
By choosing $x(\tau,\sigma) = v t + \xi(\sigma)$  in \eomcon~ one obtains in terms of the worldsheet conserved quantity, $\pi_\xi$, 
\eqn\momenta{ 
\xi^\prime = \pm  \pi_\xi {1 \over f \sqrt{h}} \sqrt{ {f-v^2  \over   
  f h-\pi_\xi^2 } }\,.
}
This  expression is cross-checked with the generic, diagonal metric, solution of \CaceresAS.
The expression \momenta~  is much like the equivalent one in ${\cal N}=4$~SYM  \GubserBZ.
Demanding that any piece of the string has a subluminal velocity amounts to fix the sign of the 
squared root in \momenta~ at certain critical point $u_{cr}^3= u_T^3/(1-v^2)\,.$ This leads to the fixation of the constant of motion
\eqn\pis{
\pi_\xi= \left({u_T\over R}\right)^{3/2}  {v\over\sqrt{1-v^2} }
\,.
} 
The main feature of the latter expression w.r.t. the ${\cal N}=4$~ case is the equal behaviour in terms of the velocity.  This probably signals an universal type of behaviour, independently whether the model is derived from a supersymmetric one or not. 
Inserting \pis~in \momenta~ and using \finite~ leads to a remarkably simple relation
\eqn\trajectory{
x(\sigma,\tau)= v \tau  
+{1 \over 3} \sqrt{{R^3 \over u_T}} v  \left\{ \sqrt{3} 
{\rm tan}^{-1} \left( {u_T+2 \sigma\over \sqrt{3} u_T} \right) 
+ \log\left( { \sqrt{u_T^2 + u_T \sigma + \sigma^2}\over u_T -\sigma } \right)
\right\}\,.
}
The shape of \trajectory~ in the $x-u$ plane at a given time is a monotonic bended curve. One could ask whether this bending signals the interaction of the quark with the reservoir of gluons. To answer this we look at the density of energy and the $x$-component momentum, \lags~does not depend explicitly neither on $t$ nor in $x$, obtained from the conserved currents,
\eqn\ccurrentsD{
\pmatrix{
\pi_t^0 \cr
\pi_x^0\cr}
= - {1\over f \sqrt{-g}}
\pmatrix{
f \left[1+h f  \left(x^\prime\right)^2\right]  \cr
- v  \cr}\,,\quad
\pmatrix{
\pi_t^1 \cr
\pi_x^1\cr}=
 {1\over \sqrt{-g}}
\pmatrix{
v f  h x^\prime \cr
-f h  x^\prime\cr}\,.
}
Inserting the configuration \trajectory~in \ccurrentsD~we find that, as in the ${\cal N}=4$ case, the two currents $\pi^1_x,\pi^1_t$ are constant along the string 
\eqn\flow{
v \pi^1_x=-\pi^1_t =  \left({ u_T \over R }\right)^{3/2} 
{v^2\over \sqrt{1-v^2}}\,.
}
The features of \flow~are: {\sl i)} it vanishes for $v\to0$ indicating that no force is needed if the quark is kept at rest w.r.t. the flow.  {\sl ii)} If we pull the quark with constant velocity, transmitting to it energy at the ratio $\pi^1_t $, this energy goes along the string were it is dissipated into the media. The fraction of energy supplied to the surrounded thermal bath at a given point along the string is constant.
If one calculates the total of energy supply to the system, $\int_{u_T}^\infty d\sigma \pi^1_t $, turns to be infinite. 
This is clearly understood if we bear in mind that we deal with infinitely massive quarks, and hence an infinite amount of energy is needed in order to move it.
As a matter of fact, without the addition of probe-branes \KarchSH~ to hold the quark at finite distance from the horizon, the only stable solution is to set the quark at infinity.  If we were dealing with the latter set up, finite quark masses, the lost of energy would be finite at any point, and the total energy needed to pull the quark would be 
proportional to its mass (location along the radial coordinate).

  
\subsec{Confined phase}

We comment briefly on the setup at zero temperature. The expectations are the following: at zero temperature and a strong coupling constant we know that quarks are confined, they are presented in singlets of color. It is mandatory then, that an embedding as the previous one, a single quark with velocity $v$,  fails to be stable.
The lagrangian density in the static gauge reads
\eqn\lagdos{
{\cal L} = -{1\over 2 \pi \alpha} \sqrt{ {\left(u^\prime\right)^2 \over f} \left[1- \left( \dot{x}\right)^2 \right]
+h \left(x^\prime\right)^2}\,,
}
that leads to the following e.o.m.
\eqn\eomdos{
 {\left(u^\prime\right)^2 \over f} \partial_\tau\left({\dot{x}\over \sqrt{-g} }\right) - 
\partial_\sigma \left( {h x^\prime \over \sqrt{-g} }\right) = 0\,.
}
As in the parallel previous case, \lagdos~does not depend explicitly on $x$ and there is a conserved worldsheet current, $\pi_\xi$. Solving \eomdos~in terms of these we have
\eqn\xidos{
\xi^\prime = \pm  \pi_\xi {1\over \sqrt{f h}} \sqrt{ {1-v^2  \over   
 h-\pi_\xi^2 } }\,.
}
Notice that in the limit $f\to 1$ \lagdos-\xidos~reduce to the equivalent expressions of the previous section as it should be.

Let's inspect \xidos. The functions $f$
and $h$ are strictly positive in their domains. Hence there is no way of fixing the conserved quantity $\pi_\xi$ as was done in \momenta: the {\sl critical} value for the radial coordinate is $u_{cr}=R\pi_\xi^{2/3}$, thus as we increase the momenta the penetration of the string in bulk also increases, but at  finite $u_{cr}$ the velocity of the string turns to be supraluminical. This picture is consistent signaling an unstable configuration.

Just for ending this subsection, and it would be proven useful below, we quote  the conserved currents
\eqn\ccurrentsN{
\pmatrix{
\pi_t^0 \cr
\pi_x^0\cr}
= - {1\over f \sqrt{-g}}
\pmatrix{
1+h f  \left(x^\prime\right)^2  \cr
- v  \cr}\,,\quad
\pmatrix{
\pi_t^1 \cr
\pi_x^1\cr}=
 {1\over \sqrt{-g}}
\pmatrix{
v  h x^\prime \cr
- h  x^\prime\cr}\,,
}
that also fulfills the relation $\pi_t^1 = - v \pi_x^1\,.$

\newsec{Thermal mass}

We compute the energy (mass) for a static configuration in \finite~and check whether it satisfies 
the field theory expectations as a function of the temperature.

Using \zero~and the embedding 
$t=\tau, u =\sigma, x(\sigma,\tau)=x_0$
the mass at zero temperature is given by
\eqn\masszero{
m(T=0)=E=-\int_{u_\Lambda}^{u_m} d\sigma \pi_t^0 = -{4\over 9} R 
{\Gamma\left({2\over 3}\right)\over\Gamma\left({1\over 6}\right)}+ u_m ~_2F_1\left(-{1\over 3},{1\over 2},{2\over 3},{64\over 729} {R^3\over u_m^3}\right)\,,
}
with $u_m$ a hard cut-off that regularizes the expression.

For \finite~  one has a very simple expression
\eqn\massfinite{
m(T)=E=-\int_{u_T}^{u_m} d\sigma \pi_t^0 =   u_m-{16\pi^2 R^3\over \beta} \,.
}

Both expressions, \masszero~ and \massfinite, differ only in a constant in the asymptotic regime. This is due to the difference in integration limits.
As mentioned above the only stable solution
without introducing probe branes is to locate the quark at the boundary, $u_m\to \infty$. This makes an infinitely heavy-quark.
A possible way to renormalize \massfinite~is to substract \masszero. This will lead to the
value of the thermal mass at the boundary. The general pattern of the mass as a function of the temperature is as follows: For low-temperature 
\masszero~ is bigger than  \massfinite. At $\beta=2 \pi^{3/4} R \sqrt{\Gamma\left(1/6\right)/\Gamma\left(2/3\right)}$~both become equal and for higher temperatures they exchange their roles.
Notice that in increasing the temperature, up to its maximum value \finite~ the mass, $m(T)-m(T=0)$, decreases in accordance with the FT expectations \refs{\NeriIC,\PisarskiDB}.  This corroborates in an independent way some of the findings in \PeetersIU.

\newsec{Drag force for Quark-Antiquark system}

In this section we shall only consider a di-quark system at high temperatures. 
The configuration we shall focus on is based in the studies of \refs{\ReyIK,\MaldacenaIM}. 
The parallel 
of  \refs{\ReyIK,\MaldacenaIM} for the background \finite~has been studied thoroughly in \BrandhuberER. In the static gauge we shall use the ansatz
$
t=\tau\,, x(\sigma,\tau)=v \tau+\xi(\sigma)\,, y(\sigma)\,,
z(\sigma,\tau)=\sigma\,,
$
i.e. the velocity is perpendicular to the {\sl dipole} moment. More general ansatz can be studied \refs{\ArgyresVS,\AvramisEM} but
in view of our results we do not expect qualitative changes w.r.t. them. 

First we consider some general features of \finite~w.r.t. this embedding \ChernicoffHI .
The lagrangian density reads
\eqn\lagqq{
{\cal L}= \left( \left[1-{v^2\over f} \right] \left[1+ f h \left(y^\prime\right)^2\right]+f h \left(\xi^\prime\right)^2\right)^{1/2}\,.
}
From the e.o.m. we obtain the trajectory in terms of worldsheet associated currents 
\eqn\pisqq{
\xi^\prime = \pm \pi_\xi  {f-v^2 \over f \sqrt{ h (f-v^2)(f h -\pi_\xi^2)-f h \pi_y^2 }}\,,
\quad
y^\prime = \pm \pi_y  {1 \over \sqrt{ h (f-v^2)(f h -\pi_\xi^2)-f h \pi_y^2 }}\,.
 }
 As in \momenta, due to the signature of the factor inside the squared root,  exists a condition on the minimal value of the radial coordinate. In the case of a string hanging from a quark was $u\ge u_{\rm cr}$.
Using \pisqq~ we find that 
$
u_{\rm min} \ge u_{\rm {cr}}
$
signaling that the string stretching between quarks can not penetrate up to the horizon. The equality,
$
u_{\rm min} = u_{\rm cr}\,,
$
only holds if $\pi_y=0.$
As a last remark we noticed that the only possible string configuration has the string parallel to the $z$-axis: the condition for a derivable curve at $r_{\rm min}$ imposes  
$dx/dy\vert_{r_{\rm min} }\to \infty$ that only is posible if $\pi_x \to 0$. As a consequence, $x^\prime=0$ and the string moves with constant velocity and with a shape parallel to the $z$-axis.

After these remarks we
restrict the above ansatz to the case $\xi(\sigma)=0.$ 
In order to obtain the effect of a possible drag force one could boost the system from the plasma reference frame to the 
di-quark system one: $x\to {x-vt\over \sqrt{1-v^2}}\,,\quad t\to {t-vx\over \sqrt{1-v^2}}\,.$ Even if it is customary to do so one should bear in mind that the experiment is in the plasma frame, and we stick to it in the remainder.
Whether we perform the boost or not the profile of a string with velocity $v$ along the $x$-axis and extended in the radial coordinate $z$ is unaltered because the boost is perpendicular to its shape
\eqn\profileqq{ 
y^\prime= \pm R^3 \pi_y  {1 \over \sqrt{\left(u_T^3-u^3\right)\left(u_T^3+R^3 \pi_y^2 - (1-v^2) u^3\right)} }\,,
\quad
u_{\rm min}^3= {u_T^3+R^3 \pi_y^2\over 1-v^2}\,.
}
The end points separation of the strings leads at a distance $L$ one from each other
\eqn\lnqq{
L\left(\pi_y,v\right)= 2 \int_{u_{\rm min}}^\infty du\, y^\prime\,.
}
To perform the integrations we follow the prescription in \ChernicoffHI~and change variables to  $h\equiv 1- \left(u_T/u\right)^3$. In addition we also make use of the rescaled momenta
$f_y^2\equiv \pi_y^2 \left(R/u_T\right)^3.$ With these ingredients the integration domain becomes
$[ (f_y^2+v^2)/(1+f_y^2),1]$. After some algebra, and using \finite, \lnqq ~reads
\eqn\newl{
L(f_y,v)= {f_y\over 2 \pi T} \int_{h_{\rm min}}^1 dh {1\over (1-h)^{1/3}\sqrt{ h\left( f_y^2 (h-1) + h -v^2\right)}}\,.
}  
This integral can be casted in terms of hypergeometric  function
\eqn\meijer{
L= i f_y \left[{\sqrt{\pi}\over \sqrt{f_y^2+v^2}} {\Gamma\left({2\over 3}\right) \over \Gamma\left( {7\over 6}\right)} ~_2F_1\left({1\over 2},  {1\over 2},{7\over 2},{1+f_y^2\over v^2+f_y^2} \right)
 - {\pi\over \sqrt{f_y^2+1}} 
~_2F_1\left({1\over 2},  {1\over 3},1,{v^2+f_y^2\over 1+f_y^2} \right)\right]
.}
Despite its appearance its domain remains in the reals.
\medskip
\ifig\fnewl{(r.h.s) Diquark system separation in unit of $1/(2\pi T)\,.$ From top to bottom (blue) curves displays the values $v=0, 0.45, 0.7$ and $0.99$. The (green) thicker vertical curve is the fit to the function $L_{\rm max}(f_y,v)$. (lhs) depicts, with the data in boxes, the maximum distance as a function of the velocity. The green curve is a quadratic fit. The triangles are the screening length.}{
\epsfxsize 2.2 in\epsfbox{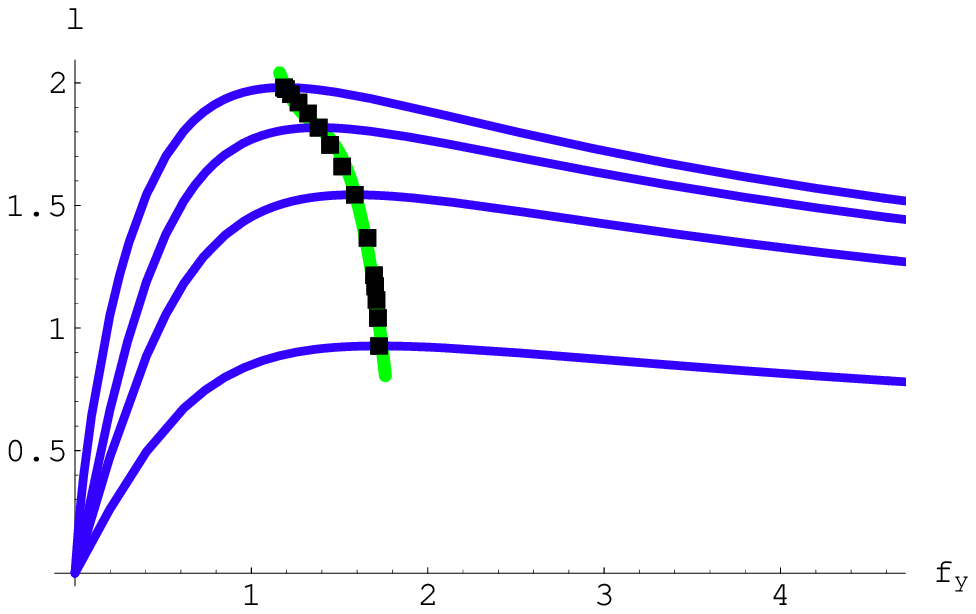}\quad\quad\quad\quad
\epsfxsize 2.2 in\epsfbox{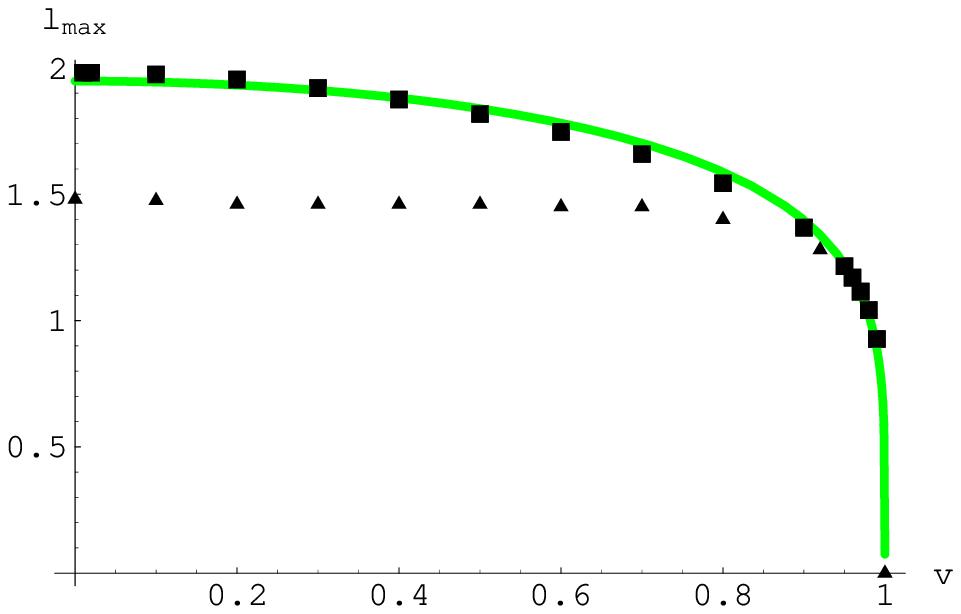}
}

In the l.h.s of \fnewl~ we have plotted the quantity $l\equiv  2 \pi T L$ as a function of the external force $f_y$.
The results are much like the same as in the supersymmetric case \ChernicoffHI~ and also is in accordance with the static one \refs{\ReyBQ,\BrandhuberBS}: the separation between quarks is not a monotonic function of the external force. As a consequence for a given velocity $v$ there is a maximum separation, $L_{\rm max}(f_y,v)$. This
function can be written analytically in terms of hypergeometric functions, but its actual expression is not very illuminating. Instead of presenting it we have added to the figure a few representative points, and a cubic polynomial fit to them. In the r.h.s of \fnewl~we have evaluated the maximum separation between quarks as a function of the velocity. Points belong to the same set as in the r.h.s figure and the green curve is a fit. The best $\chi$-squared is obtained with the function
$$
L_{\rm max}(v)\approx {1\over 2\pi T} 1.9 \left(1- v^2\right)^{1/5}\,.
$$
For $v>0.98$ there is a pronounced drop in the curve and it vanishes for $v=1$.

\lref\GrossBR{
  D.~J.~Gross, R.~D.~Pisarski and L.~G.~Yaffe,
  ``QCD And Instantons At Finite Temperature,''
  Rev.\ Mod.\ Phys.\  {\bf 53}, 43 (1981).
}

\lref\LiuNN{
  H.~Liu, K.~Rajagopal and U.~A.~Wiedemann,
  ``An AdS/CFT calculation of screening in a hot wind,''
  arXiv:hep-ph/0607062.
}

We turn now to the evaluation of the string energy in the plasma rest frame.
Using \profileqq~the density lagragean \lagqq~ reduces to
\eqn\densityqq{
{\cal L}=  {(1-v^2) u^3 -u_T^3 \over \sqrt{ \left(u^3-u_T^3\right) \left((1-v^2)u^3-u_T^3-R^3\pi_y^3 \right) }}\,,
}
and the string action reads
\eqn\enrqq{
{\cal S}\left(\pi_y,v\right)=  {1\over \pi \alpha^\prime {\cal T}} \int_{-{\cal T}/2}^{{\cal T}/2} \int_{u_{min}}^\infty 
dt\, dr\, {\cal L}\,.
} 
The above expression, \enrqq, diverges  and needs to be regularized.  This is more easily seen by looking the behaviour of \densityqq~ at the boundary: $\sqrt{1-v^2}$, which after the 
worldsheet integration per boundary unit time of \enrqq~becomes linear divergent. One possibility to  cure this is to subtract the action of a pair of quarks stretching from the boundary up to the horizon
\eqn\ensuqq{
{\cal S}\left(\pi_y,v\right)=  {1\over \pi \alpha^\prime {\cal T}} \int_{-{\cal T}/2}^{{\cal T}/2} 
dt\, \left(\int_{u_{min}}^\infty  dr\, {\cal L}-\int_{u_{T}}^\infty  dr\, {\cal L}\right)\,.
} 
This quantity is related with the expectation value of the Wilson loop operator. Even if it is an interesting quantity at zero temperature, in pure gauge theory at finite temperature  is the Polyakov-loop the order parameter which signals whether the system is confined or not \GrossBR.  For this reason instead of evaluating \ensuqq~we compute the energy stored in the di-quark system.

Using \enrqq~one can already compute its associated conserved canonical momentum densities 
\eqn\momdi{
\pmatrix{
\pi_t^0 \cr
\pi_x^0\cr
\pi_x^0\cr}
= {1\over \sqrt{-g}}
\pmatrix{
-  1- f h \left(y^\prime\right)^2 \cr
v  [{1\over f} +h\left(y^\prime\right)^2] \cr
0 \cr}\,,\quad
\pmatrix{
\pi_t^1 \cr
\pi_x^1\cr
\pi_u^1\cr}=
 {1\over \sqrt{-g}}
\pmatrix{
0 \cr
0\cr
(v^2-f)h\left(y^\prime\right)^2 \cr}\,.
}
As one can see the component $\pi_x^1$ vanishes signaling that, contrary to the single quark configuration, the di-quark experiences no force by the presence of the gluon bath in the direction on its movement, in that sense one can recover the experimental data that signals the dynamics of quarks inside an almost perfect fluid. Notice
eventhought that exists a force  in the $y$-direction.This behaviour is reminiscence to the image of the Cooper pairs in super-conductivity: one electron can feel a drift while a pair does not. 

As already happened with \enrqq~the energy of the system will need to be renormalized. We shall chose the standard method subtracting the bare mass of a pair of quarks, but other scheme, as for instance subtracting the corresponding energy obtained from \zero, as was done in sec. 3, will do the same job as  far as it has the same asymptotic behaviour. 
The final result reads
\eqn\ensub{ E(f_y,v)= 
{u_T\over 3} \left(\int_{h_{min}}^1\!\!\! dh {h^{1/2} \over (1-h)^{4/3}\sqrt{ f_y^2 (h-1) + h -v^2}} -
\int_0^1\! \!dh {h^{1/2} \over (1-h)^{4/3}\sqrt{1 -v^2}}
\right).}
In the asymptotic regime this expression vanishes. More over due to an infrared singularity \ensub~is not uniquely determined \ChernicoffHI.
\ifig\evsl{(r.h.s.) Energy as a function of the quark separation for different velocities. From top to bottom $v= 0.8, 0.7, 0.4$ and $0$. (l.h.s.) The maximum energy \ensub~as a function of the velocity. Data points are take from the r.h.s. plot. The curve is a quadratic fit to the point with $v< 0.6$.}{
\epsfxsize 2.2 in\epsfbox{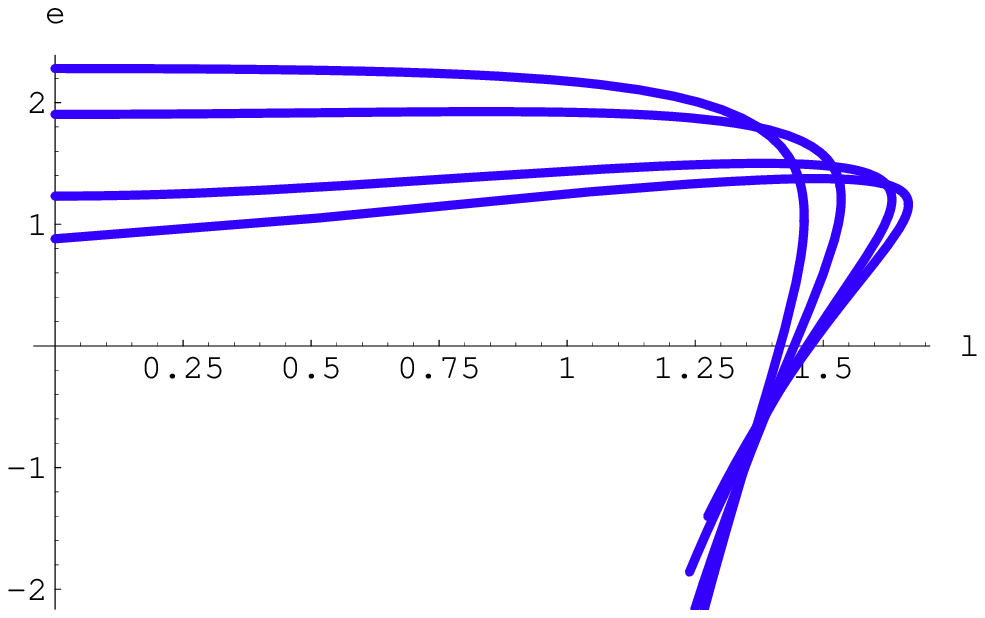}
\quad\quad\quad\quad
\epsfxsize 2.2 in\epsfbox{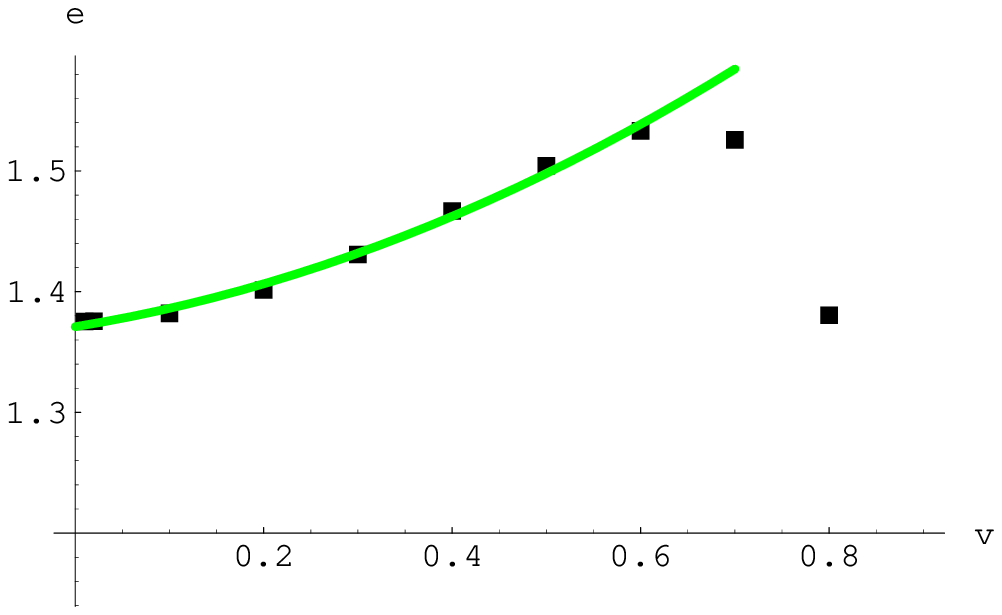}
}
We have plotted in the l.h.s. of \evsl~ a normalized energy, $e= 3 E/( u_T)$, vs. the separation of a quark pair at different velocities, that turns to be a bivaluated function. Bound states are allowed to exist only in the region were $e< 0$. If $e>0$ the energy of a pair of free quarks is less than the energy of the bound state and chiral symmetry restoration has occurred. The separation between quarks where this phenomenon happens is defined as the screening length ($L_*$) \refs{\LiuNN,\ChernicoffHI}: as increasing the distance between quarks gluons start to screen the quark interaction until  it becomes negligible and deconfinement has occurred. Notice that if the system has some velocity $v$ the screening length diminishes very smoothly. In the static case one finds $L_*\sim 1.34/(2 \pi T)$, slightly lower that the ${\cal N}=4$ case. The relevant feature of \fnewl~r.h.s.  is that the screening length is almost a constant factor as a function of the velocity for $v<0.9$ in agreement with the conclusions in \refs{\LiuNN,\CaceresTA} where it was spelled out that the main velocity dependence comes from the boost action. In this region $L_*$~never reaches the maximum possible separation between quarks. i.e. matter deconfinement occurs at shorter distances than naively expected, a factor 
$\sim 3/2$ lower than the expected. From $v>0.9$ on, $v\to1$, the screening length, $L_*$, is given by the maximum separation between quarks and goes to zero.

Inspecting more carefully \evsl~l.h.s. the salient points are: {\sl i)} For each separation between quarks we find two possible energetic states: a lower energy position, that is the stable one, and an upper that indicates some metastable state. For $v>>$ and for high energies the curve is almost flat with a very large energetic gap between the two possible states given a di-quark separation. This indicates that the metastable states are very unstable. For all the curves, the rhs part of the plot, very short separations, can only be reached for $f_y\to\infty$. {\sl ii)} For a fixed energy, there are also two possible positions, the one with the minimum separation between quarks correspond to the set of metastable states. Although not very intuitive quarks do not lie as near as they can from each other without decaying. 
{\sl iii)} As we increase the velocity the curve moves to the right and upwards. One should wonder about the different shape and velocity behaviour for the cases of medium-high velocities in \evsl~w.r.t. the equivalent ones in \ChernicoffHI. We have checked that this difference comes from the boost, if we go to the di-quark rest frame the qualitative behaviour between the two models agree.

In the r.h.s of \evsl~we have plotted the maximum energy of the di-quark system at a given velocity. Up to $v\sim 0.9$ this maximum energy is reach at the maximum separation point between quarks and is an  increasing function. At this precise point, $v\sim 0.9$, the maximum of energy is not longer found at this distance and we find a decreasing function that diverges at $v\to 1$.

\newsec{Summary}
We have explored the drag force for a single and a di-quark system in a dual model to large-N QCD.  Particularly relevant is that the gross features we find in this model are in qualitative agreement with those derived in others models obtained from ${\cal N}=4$. This signals that the main physical aspects are not sensible to the initial model we depart at the very beginning, before breaking supersymmetry incorporating temperature, is supersymmetric or not.

Our results for the single quark are parallel to previous studies.
The behaviour of the thermal mass as a function of the temperature is in agreement with the field theory expectations.

In addition the di-quark system shares most of the peculiarities of other systems already studied in the literature. The main relevant difference coming from the {\sl almost} constancy, up to velocities $v< 0.9$, of the screening length as a function of the velocity and its relative smallness in from of the maximum allowed distance between quarks.


\bigskip

{\bf Acknowledgments}

\smallskip

I would like to thank Ll. Ametller for conversations and a careful reading of the manuscript.
Also to A. Cotrone and J. M. Pons for discussions. To D Zeng for pointing out an error in an earlier version.

This research
was supported in part by Generalitat de Catalunya BE2006 ,
the European Community's
Human Potential Programme under contract MRTN-CT-2004-005104
`Constituents, fundamental forces and symmetries of the universe' and by the Programme RyC.
I also thank the MIT LNS laboratory where part of this work was done.

\listrefs

\bye